\documentclass[twocolumn,showpacs,preprintnumbers,amsmath,amssymb]{revtex4}
\usepackage{amsfonts}
\usepackage{mathrsfs}
\usepackage{graphicx}% Include figure files
\usepackage{dcolumn}% Align table columns on decimal point
\usepackage{bm}% bold math

\begin{document}
\title{Particle Entanglement in Rotating Gases}
\author{Zhao Liu}%
 \email{liuzhaophys@aphy.iphy.ac.cn}
\author{Heng Fan}
%\email{hfan@aphy.iphy.ac.cn}
%\author{}
%\author{}
\affiliation{%
Institute of Physics, Chinese Academy of Sciences, Beijing 100190,
China
}%
\date{\today}% It is always \today, today,

\begin{abstract}
In this paper, we investigate the particle entanglement in 2D
weakly-interacting rotating Bose and Fermi gases. We find that both
particle localization and vortex localization can be indicated by
particle entanglement. We also use particle entanglement to show the
occurrence of edge reconstruction of rotating fermions. The
different properties of condensate phase and vortex liquid phase of
bosons can be reflected by particle entanglement and in vortex
liquid phase we construct the same trial wave function with that in
[Phys. Rev. Lett. {\bf 87}, 120405 (2001)] from the viewpoint of
entanglement to relate the ground state with quantum Hall state.
Finally, the relation between particle entanglement and interaction
strength is studied.
\end{abstract}

%{\bf dynamical critical
%phenomena, spin chain models, order-disorder transformations }
\maketitle

%\section{\label{sec:level1}First-level heading:\protect\\ The line
%break was forced \lowercase{via} \textbackslash\textbackslash}

{\label{sec:level1}}
\section{Introduction}
Quantum entanglement, which is considered to be the most
non-classical phenomenon in the quantum world, has been attracting
much attention in last decades. On the one hand, it has been
identified as a key resource in many aspects of quantum information
theory, such as quantum teleportation, quantum key distribution and
quantum computation \cite{book}. On the other hand, the concept of
entanglement is proven to be useful in condensed matter systems
\cite{Vidal}, such as in spin chain, Bose-Einstein condensate and
electronic quantum Hall effect.

As a typical condensed matter system, the two-dimensional (2D)
rotating atom gas is a very interesting field because it opens a
door to many important quantum phenomena \cite{Bloch,ALF}. Many
excellent experimental and theoretical papers focus on the formation
and melt of vortex lattice \cite{BR}, many-body energy spectrum
\cite{GFB,RN}, its analogy to quantum Hall effect of electrons in a
magnetic field \cite{NNJ,NRa,NRb,NRc,Viefer,KO}, particle
localization and vortex localization \cite{IR,SMRa,SMRb,LOB}, the
comparison between the results of exact quantum numerical solution
and mean-field theory \cite{DNMJ} and so on. However, the
entanglement in this system has not been investigated as extensively
as in other condensed matter systems, for example in spin chain
systems.

In one of our previous work, we use some quantum information
concepts, such as fidelity susceptibility and particle entanglement
to study the properties of 2D weakly-interacting rotating
Bose-Einstein condensate trapped in a harmonic potential \cite{Liu}.
We find that the single-particle entanglement, defined as the von
Neumann entropy of the single-particle reduced density operator, can
indicate some angular momentum of real ground states and reflect the
properties of condensed phase (single-vortex state) and vortex
liquid phase. This indicates that in 2D rotating atom (both boson
and fermion) gas, some other important properties of this rotating
system may be shown by entanglement.

In this paper, we mainly investigate the particle entanglement of 2D
weakly-interacting rotating atom gas trapped in a harmonic potential
in disk geometry. In order to compare the results of bosons and
fermions, we choose repulsive Coulomb interaction. For a fixed
particle number $N$, we calculate the single-particle entanglement
$S_{1}$ of the ground states in a series of subspaces of fixed
angular momentum $L$. We find that both particle localization and
vortex localization can be reflected by the oscillation of $S_{1}$
with the increase of $L$. This oscillation is universal for bosons
and fermions. Then we study the single-particle entanglement of the
ground states in subspaces whose angular momentum are special
functions of $N$. For the subspace $L=N$, through the comparison
between the results of bosons and fermions, we find that the edge
reconstruction of rotating fermions can be indicated by the
single-particle entanglement. For the subspace
$L=(N-\bar{N})(N+\bar{N}-k)/k$ with $k\geq1$ an integer and
$\bar{N}$ the smallest non-negative integer making $N-\bar{N}$ to be
divisible by $k$, we relate its subspace ground state with quantum
Hall state from the viewpoint of entanglement and construct the same
trial wave function as that in Ref.\cite{NNJ}. Finally, we study
rotating bosons trapped in a quadratic plus quartic trap interacting
with contact potential, both attractive and repulsive. We find both
single-particle and two-particle entanglement of the ground state in
subspace $L=N$ and $L=N(N-1)$ increase with the interaction strength
for fixed $N$. However, when the interaction is attractive, the
single-particle entanglement of the ground state in subspace $L=N$
does not decay with $N$, verifying that attractive bosons do not
condensate \cite{NKW}.
\section{Model}
We focus our attention on the 2D rotating spinless Bose and
polarized Fermi atom gases. In rotating reference, the Hamiltonian
of $N$ atoms is as follows:
\begin{eqnarray}
\mathcal
{H}=\sum_{i=1}^{N}\Big\{-\frac{\hbar^{2}}{2m}\nabla_{i}^{2}+\mathcal
{V}(r_{i})-L_{z,i}\Omega\Big\}+\sum_{i<j=1}^{N}\mathcal
{U}(\textbf{r}_{i},\textbf{r}_{j}),\nonumber\\
 \label{e1}
\end{eqnarray}
where $\mathcal {V}(r)$ is the trap potential, $-L_{z}\Omega$ is the
rotating energy with rotation frequency $\Omega$ and $\mathcal {U}$
is the interaction energy. $\mathcal {U}>(<)0$ represents repulsive
(attractive) interaction. Typically, $\mathcal
{V}(r)=\frac{1}{2}m\omega^{2}r^{2}$ is a harmonic potential.
$\mathcal {U}$ can be either a short-range contact interaction
$\delta(\textbf{r}_{i}-\textbf{r}_{j})$ or a long-range Coulomb
interaction $1/|\textbf{r}_{i}-\textbf{r}_{j}|$.

In the weakly-interacting limit, we can use single-particle lowest
Landau level (LLL) wave function $\varphi_{l}(z)=\frac{1}{\sqrt{\pi
l!}}z^{l}e^{-|z|^{2}/2}$, where
$z=(x+\textrm{i}y)\sqrt{m\omega/\hbar}$ is the dimensionless
position of a particle, to make second-quantization of
Eq.(\ref{e1}). The obtained Hamiltonian in the subspace of fixed
angular momentum $L_{z}=L\hbar$ takes the form:
\begin{eqnarray}
\mathcal
{H}_{L}=(L+N)\hbar\omega-L\hbar\Omega+U_{0}\sum_{i,j,k,l=0}^{L}U_{i,j,k,l}a_{i}^{\dag}a_{j}^{\dag}a_{k}a_{l},\label{e2}
\end{eqnarray}
where $|U_{0}|$ represents the strength of interaction and
\begin{eqnarray}
U_{i,j,k,l}=\int\!\!\!\!\int
dz_{1}dz_{2}\varphi_{i}^{*}(z_{1})\varphi_{j}^{*}(z_{2})\mathcal
{U}(z_{1},z_{2})\varphi_{k}(z_{2})\varphi_{l}(z_{1}).\nonumber
\end{eqnarray}
For contact interaction, the expression is simple:
$U_{i,j,k,l}=\frac{1}{2^{i+j}}\frac{(i+j)!}{\sqrt{i!j!k!l!}}\delta_{i+j,k+l}$.
But for Coulomb interaction, the expression is a little complicated
\cite{tsiper}:
\begin{eqnarray}
U_{i,j,k,l}=\delta_{i+j,k+l}\sqrt{\frac{i!k!}{j!l!}}\frac{\Gamma(i+j+3/2)}{2^{i+j}}\nonumber\\
%\frac{\textrm{\Gamma}(i+j+3/2)}{2^{i+j}}
\times(A_{lj}^{i-l}B_{jl}^{i-l}+A_{jl}^{i-l}B_{lj}^{i-l}),\nonumber
\end{eqnarray}
where
\begin{eqnarray}
A_{rs}^{t}&=&\sum_{i=0}^{r}\frac{r!}{i!(r-i)!}\frac{\Gamma(i+1/2)\Gamma(i+t+1/2)}{(i+t)!\Gamma(i+s+t+3/2)},\nonumber\\
B_{rs}^{t}&=&\sum_{i=0}^{r}\frac{r!}{i!(r-i)!}\frac{\Gamma(i+1/2)\Gamma(i+t+1/2)}{(i+t)!\Gamma(i+s+t+3/2)}
\nonumber \\
&&\times (2i+t+1/2).\nonumber
\end{eqnarray}
$\Gamma(x)=\int_{0}^{+\infty}t^{x-1}\exp(-t)dt$ is the usual Gamma
function.

After diagonalizing Eq.(\ref{e2}) numerically by Lanczos algorithm,
we obtain its subspace ground state $|\Phi_{L}\rangle$ in angular
momentum Fock representation. Here we consider entanglement between
particles \cite{Haque1} (In Ref.\cite{AGD}, the entanglement between
angular momentum orbits is investigated to obtain the topological
entropy in rotating Bose-Einstein condensate). We define the
single-particle reduced density operator of $|\Phi_{L}\rangle$ as
$(\rho_{1})_{ij}=\frac{1}{N}\langle
a_{j}^{\dag}a_{i}\rangle_{L}=\frac{1}{N}\delta_{i,j}\langle
a_{i}^{\dag}a_{i}\rangle_{L}$, with
$\langle\cdot\rangle_{L}\equiv\langle\Phi_{L}|\cdot|\Phi_{L}\rangle$.
The single-particle entanglement of $|\Phi_{L}\rangle$ is just the
von-Neumann entropy of $\rho_{1}$:
\begin{eqnarray}
S_{1}=-\textrm{Tr}(\rho_{1}\ln\rho_{1})=\ln
N-\frac{1}{N}\sum_{i=0}^{L}\langle
a_{i}^{\dag}a_{i}\rangle_{L}\ln\langle
a_{i}^{\dag}a_{i}\rangle_{L}.\nonumber
\end{eqnarray}
For bosons, it's possible that $\langle
a_{i}^{\dag}a_{i}\rangle_{L}=N$ for one $i$ and $\langle
a_{j}^{\dag}a_{j}\rangle_{L}=0$ for $j\neq i$. Therefore
$S_{1}\geq0$. For fermions, because $\langle
a_{i}^{\dag}a_{i}\rangle_{L}\leq1$, we can find that $S_{1}\geq\ln
N$. Moreover, considering the dimension of single-particle Hilbert
space is $L$, we have $S_{1}\leq\ln L$ for both bosons and fermions.

Similarly, we can define the two-particle reduced density operator
of $|\Phi_{L}\rangle$ as $(\rho_{2})_{ij,kl}=\frac{1}{N(N-1)}\langle
a_{k}^{\dag}a_{l}^{\dag}a_{j}a_{i}\rangle_{L}\delta_{i+j,k+l}$ and
its two-particle entanglement as
$S_{2}=-\textrm{Tr}(\rho_{2}\ln\rho_{2})$.

\section{Single-particle entanglement}
In this section we consider single-particle entanglement of both
bosons and fermions interacting with each other by repulsive Coulomb
potential.

\begin{figure}
\includegraphics[height=15cm,width=\linewidth]{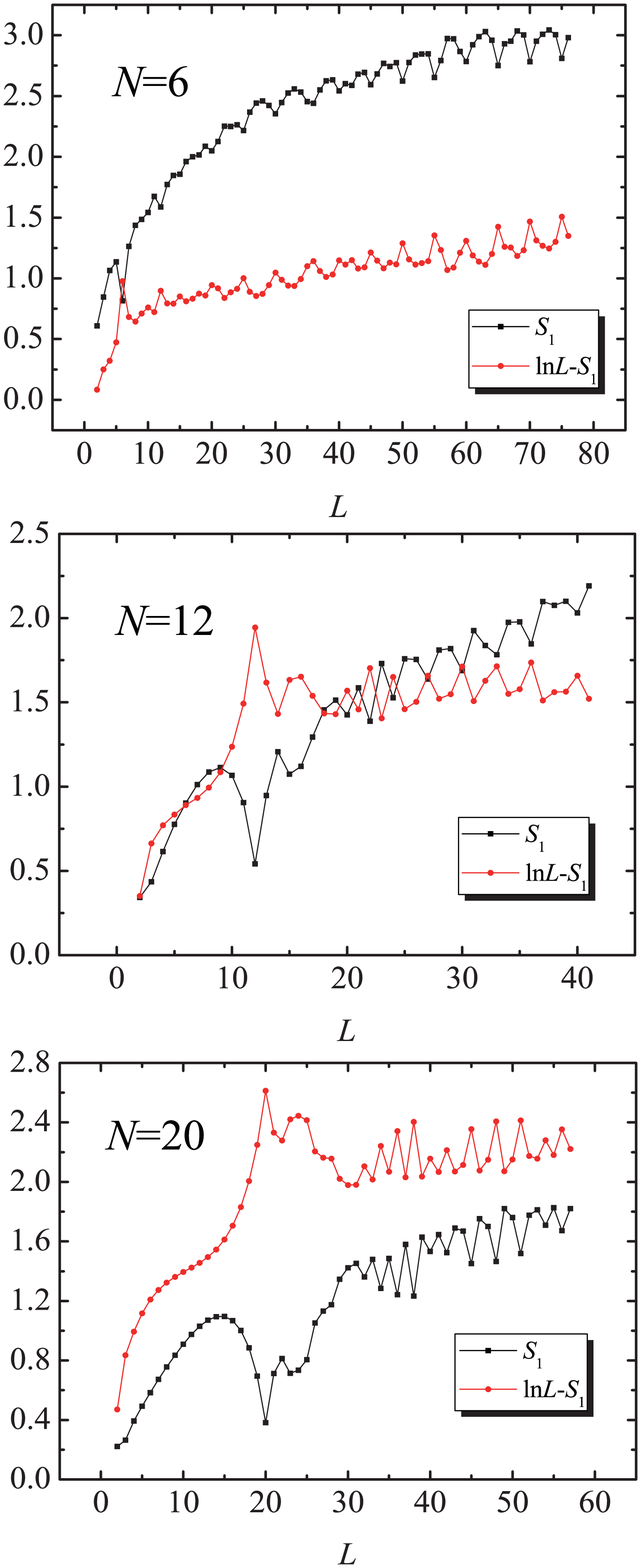}
\caption{\label{f1}(Color Online) The subspace ground state
single-particle entanglement $S_{1}$ and the difference between
$S_{1}$ and its upper bound $\ln L$ for $N$=6,12,20 bosons
interacting through repulsive Coulomb potential. The oscillations of
$S_{1}$ and $\ln L-S_{1}$ can reflect either particle localization
(for $N=6$) or vortex localization (for $N=12$ and $N=20$).}
\end{figure}

First we fix the particle number $N$ to calculate the
single-particle entanglement in a series of subspace ground states
$|\Phi_{L}\rangle$. It's known that at extreme angular momentum $L$,
a few particles ($N$ is small) in a harmonic trap localize to Wigner
molecules, independent of their bosonic or fermionic statistics.
While at moderate angular momentum, if the particle number $N$ is
much larger than the number of vortices, the localization happens in
vortices. The localization of particles (vortices) can be studied by
particle (hole) pair correlation function and can be reflected by
the regular oscillation of quantum many-body energy spectrum with
$L$ \cite{SMRa,SMRb}. As we show below, the single-particle
entanglement of $|\Phi_{L}\rangle$ also oscillates with $L$ and can
reflect particle and vortex localization very well. In Fig.\ref{f1},
we fix particle number $N$ to calculate single-particle entanglement
$S_{1}$ of the ground state of bosons in every angular momentum
subspace $L\geq 2$. With the increase of $L$, $S_{1}$ not only has
the tendency to become larger but also shows oscillation. To see the
oscillation clearer, we also plot $\ln L-S_{1}$, the difference
between $S_{1}$ and its upper bound, in Fig.\ref{f1}. For $N=6$, a
series of local maxima of $\ln L-S_{1}$ appear: $\mathcal
{L}$=6,10,12,15, 18,20,25,30,36,40,42,45,48,50,55,60,65,70,75 etc.
The angular momentum of the real ground state is
$L_{g}$=6,10,12,15,20,24,30,36,40,45,50,55,60,65,70,75 etc.
Therefore similar to the case of bosons with contact interaction,
the local maxima of $\ln L-S_{1}$ of the subspace ground states can
indicate some angular momentum of real ground state \cite{Liu}. One
can note $\mathcal {L}$ can always be written as $5k$ or $6k$ with
$k$ a positive integer. This reflects the spatial symmetry of the
particles in the subspace ground state, namely particle
localization. The series $\mathcal {L}=5k$ is associated with a
(1,5) pentagon ring structure with one particle at the center, while
the series $\mathcal {L}=6k$ is associated with a (0,6) hexagon ring
structure. When $L$ is large enough ($L\geq50$), the series
$\mathcal {L}=5k$ dominate. When $L$ is relatively small, the series
$\mathcal {L}=5k$ and $\mathcal {L}=6k$ compete with each other and
the spatial symmetry of particles can be observed by full $N$-point
correlation function of the subspace ground state \cite{LOB}. For
relative large $N$ such as $N=12$ and $N=20$, the oscillations of
$S_{1}$ and $\ln L-S_{1}$ are different from those for $N=6$. For
$N=12$, $S_{1}$ and $\ln L-S_{1}$ oscillate with a period of
$P_{L}=2$ between $L=20$ and $L=24$ corresponding to the two-vortex
state and they oscillate with a period of $P_{L}=3$ between $L=24$
and $L=36$ corresponding to the three-vortex state. For $N=20$,
$S_{1}$ and $\ln L-S_{1}$ oscillate with a period of $P_{L}=2$
between $L=32$ and $L=42$ corresponding to the two-vortex state and
they oscillate with a period of $P_{L}=3$ between $L=42$ and $L=54$
corresponding to the three-vortex state. This phenomenon reflects
vortex localization.

\begin{figure}
\includegraphics[height=15cm,width=\linewidth]{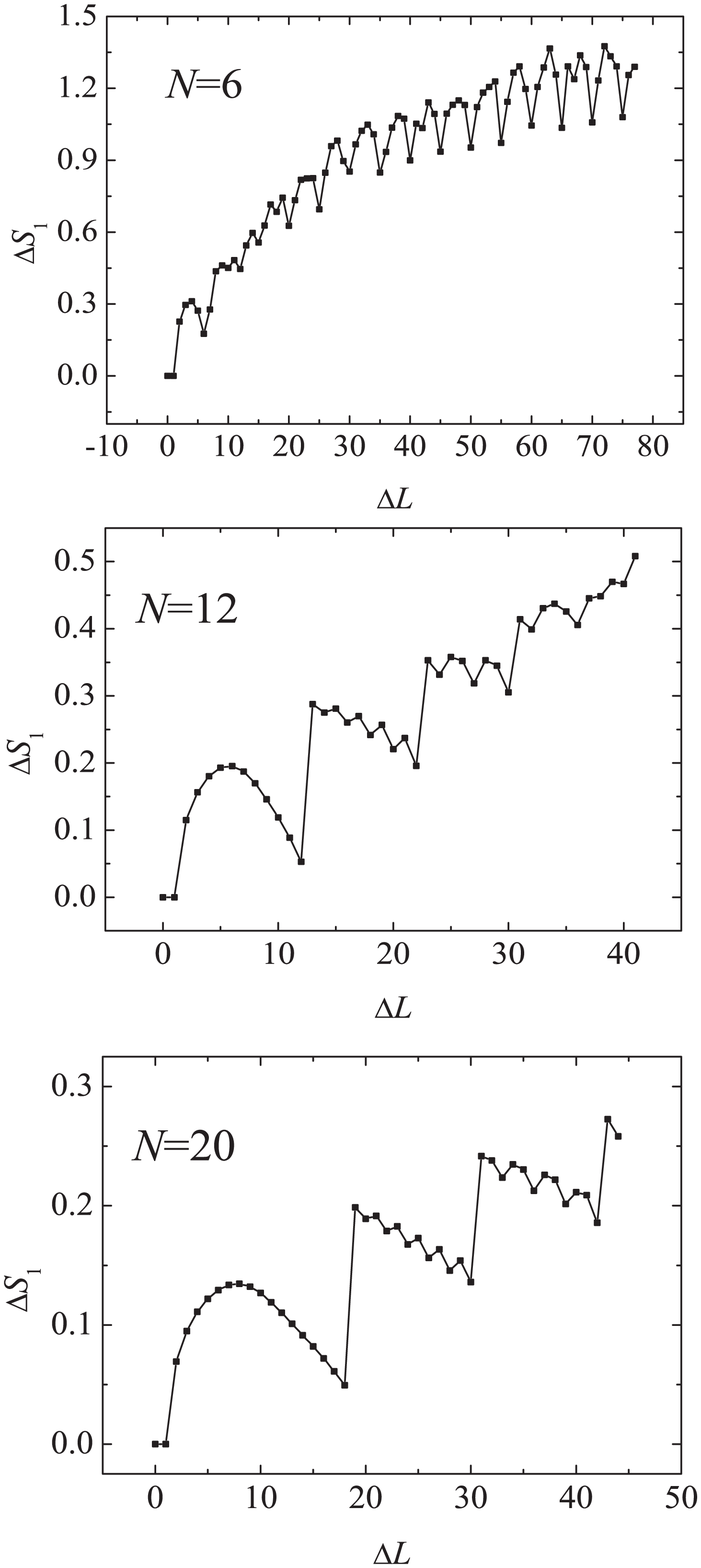}
\caption{\label{f2}The subspace ground state single-particle
entanglement $\Delta S_{1}=S_{1}-\ln N$ for $N$=6,12,20 fermions
interacting through repulsive Coulomb potential. The oscillation of
$\Delta S_{1}$ can reflect either particle localization (for $N=6$)
or vortex localization (for $N=12$ and $N=20$).}
\end{figure}

\begin{figure}
\includegraphics[height=10cm,width=\linewidth]{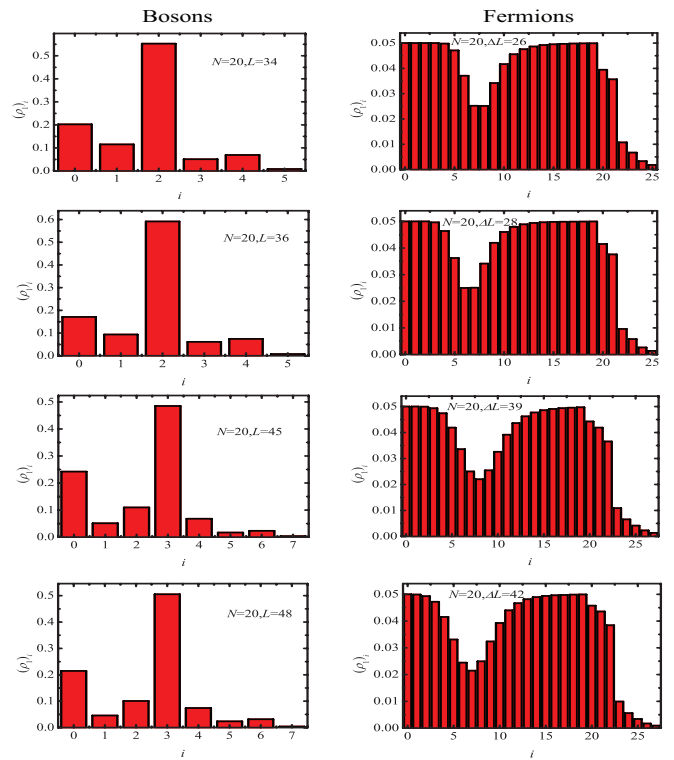}
\caption{\label{f9}(Color Online) The eigenvalues of the
single-particle reduced density operator of bosons (left column) and
fermions (right column). For bosons, the first two (three)
eigenvalues of the single-particle reduced density operator in two-
(three-) vortex state of bosons are always relatively small with the
increase of $L$. But for fermions, the smallest two (three) values
in the trough of eigenvalues of the single-particle reduced density
operator in two- (three-) vortex state of fermions move towards the
center with the increase of $L$. For example, for 20 fermions with
$\Delta L=26$ (39), the 7th and 8th (7th, 8th and 9th) eigenvalues
are in the bottom of the trough.}
\end{figure}

Due to Pauli exclusion principle, the smallest angular momentum of
$N$ fermions is $M=N(N-1)/2$, whose subspace ground state
single-particle entanglement is $\ln N$. Therefore when considering
fermions, sometimes it is convenient to use $\Delta L=L-M$ and
$\Delta S_{1}=S_{1}-\ln N$. In Fig.\ref{f2}, we fix particle number
$N$ to calculate single-particle entanglement $\Delta S_{1}$ of the
ground state of fermions in every angular momentum subspace $\Delta
L\geq0$. The oscillation of $\Delta S_{1}$ is very clear so we do
not need to plot the $\ln L-S_{1}$ as what we did for bosons. On the
one hand, there is similarity between the oscillations of subspace
ground state single-particle entanglement of bosons and fermions.
For $N=6$, the positions $\Delta \mathcal {L}$ of local minima of
fermionic $\Delta S_{1}$ are nearly the same with those $\mathcal
{L}$ for bosons, reflecting the particle localization in the
subspace ground state. For $N=12$ and $N=20$, $\Delta S_{1}$
oscillates with a period of $P_{L}=2,3$ and even 4 successively,
reflecting the vortex localization. On the other hand, there also
exists difference between boson and fermion case. For bosons, the
average angular momentum per particle $\ell_{2(3)}$ at which $S_{1}$
begins to oscillate with a period of $P_{L}=2(3)$ is nearly a
constant. When $N=12$, we have $\ell_{2}=20/12\approx1.67$ and
$\ell_{3}=24/12=2.0$. When $N=20$, we have $\ell_{2}=32/20=1.6$ and
$\ell_{3}=42/20=2.1$. This is consistent with the conclusion of
mean-field theory that two vortices form at $\ell_{2}\approx1.7$ and
three vortices form at $\ell_{3}\approx2.1$ \cite{BR}. But for
fermions, $\ell_{2(3)}$ is not a constant. When $N=12$, we have
$\ell_{2}=14/12\approx1.17$ and $\ell_{3}=24/12=2.0$. When $N=20$,
we have $\ell_{2}=20/20=1.0$ and $\ell_{3}=33/20=1.65$. Moreover,
from Fig.\ref{f9}, we can see that the change of the distribution of
eigenvalues of the single-particle reduced density operator of
bosons with the increase of $L$ is different from that of fermions.
For bosons, the first two (three) eigenvalues of the single-particle
reduced density operator in two- (three-) vortex state of bosons are
always relatively small with the increase of $L$. But for fermions,
the smallest two (three) values in the trough of eigenvalues of the
single-particle reduced density operator in two- (three-) vortex
state of fermions move towards the center with the increase of $L$.

\begin{figure}
\includegraphics[height=6cm,width=\linewidth]{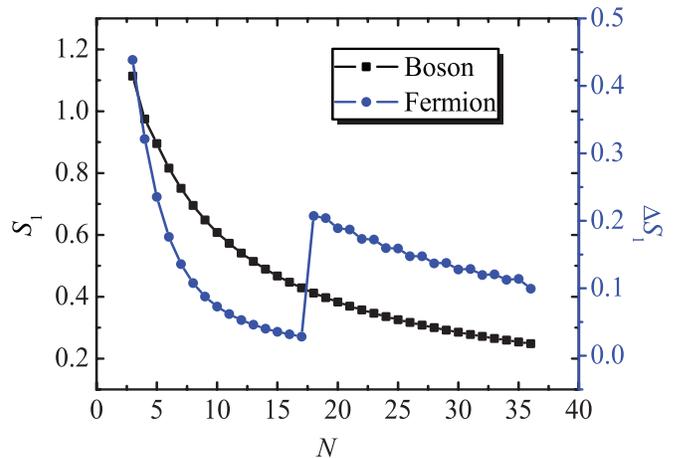}
\caption{\label{f3}(Color Online) The entanglement of the ground
state in the subspace $L=N$ for bosons and $\Delta L=N$ for
fermions. The interaction is repulsive Coulomb potential.}
\end{figure}

\begin{figure}
\includegraphics[height=4cm,width=\linewidth]{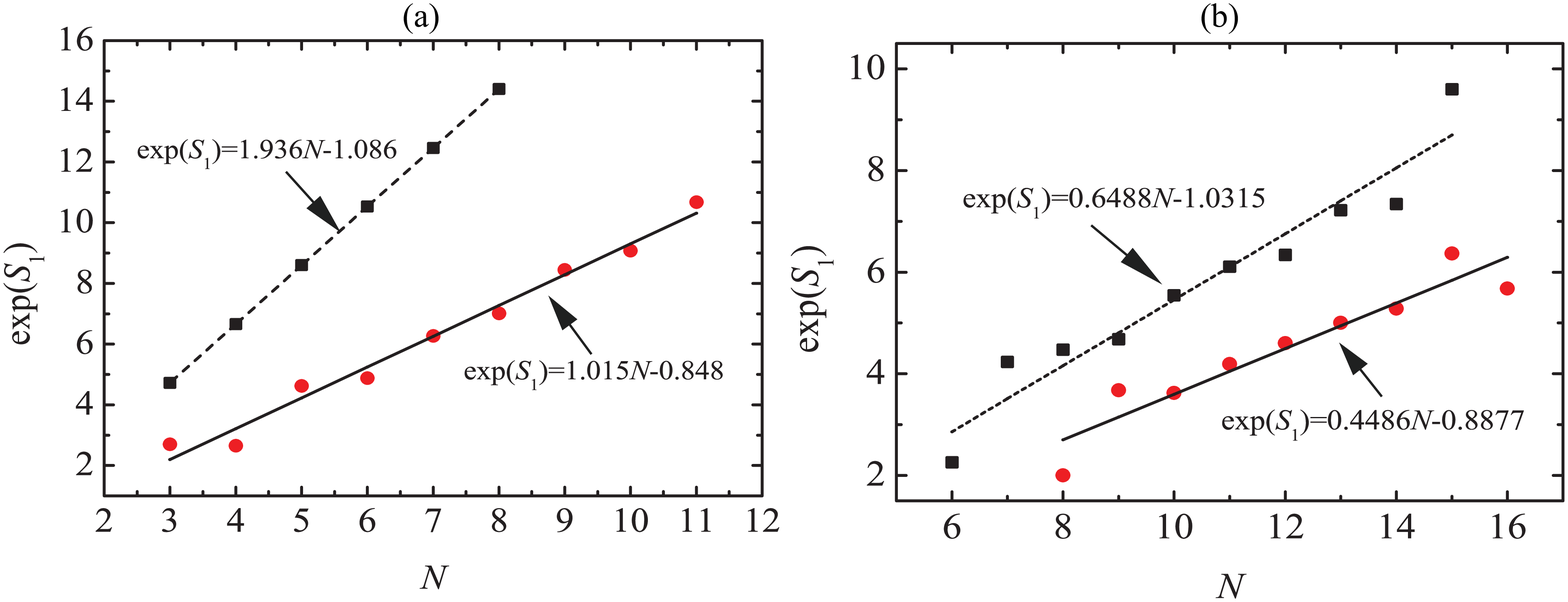}
\caption{\label{f4}(Color Online) Entanglement of bosons with
repulsive Coulomb interaction. (a) Black cubic: The entanglement of
the ground state in the subspace $L=N(N-1)$, namely $k=1$. Red
circle: The entanglement of the ground state in the subspace
$L=(N-1)^{2}/2$ for odd $N$ and $L=N(N-2)/2$ for even $N$, namely
$k=2$. (b) Black cubic: The entanglement of the ground state in the
subspace $L=(N-\bar{N})(N+\bar{N}-3)/3$, namely $k=3$. Red circle:
The entanglement of the ground state in the subspace $
L=(N-\bar{N})(N+\bar{N}-4)/4$, namely $k=4$.}
\end{figure}

Next we study the single-particle entanglement of bosons of the
ground state in some special subspaces, whose angular momentum are
functions of particle number $N$. The first subspace is $L=N$. In
Fig.\ref{f3}, we plot $S_{1}$ of the ground state in the subspace
$L=N$ for bosons. One can see that $S_{1}$ decays with $N$
monotonically, just like in the case of contact interaction
\cite{Liu}, demonstrating a property of condensate phase. The second
subspace is $L=(N-\bar{N})(N+\bar{N}-k)/k$ with $k\geq1$ an integer,
where $\bar{N}$ is the smallest non-negative integer making
$N-\bar{N}$ to be divisible by $k$. For example, when $k=1$,
$\bar{N}=0$ and when $k=2$, $\bar{N}=0$ for even $N$ and $\bar{N}=1$
for odd $N$. In Fig.\ref{f4}, we plot $S_{1}$ of the subspace ground
states corresponding to $k=1,2,3,4$. We can see for all of them
$S_{1}\approx\ln(2N/k-1)$, demonstrating a strongly-correlated
property of vortex liquid phase. This phenomenon implies from an
entanglement view that the ground states in these subspaces
$L=(N-\bar{N})(N+\bar{N}-k)/k$ may have a close relation with
quantum Hall states. Recalling that for a quantum Hall state in
spherical geometry, the single-particle entanglement is exactly
$S_{1}=\ln(N/\nu-\sigma+1)$, where $\nu$ is the filling factor
 and $\sigma$ is called shift, we can extract some information of those subspace ground states from their single-particle entanglement, although we are
 dealing with disk geometry.
Comparing with our results, we can find that the ground state of subspace $L=(N-\bar{N})(N+\bar{N}-k)/k$ is a
 quantum Hall state with filling factor $\nu=k/2=\lim_{N\rightarrow\infty}\frac{N(N-1)}{2L}$ and when $N$ is divisible by $k$ the shift of its counterpart in
 spherical geometry is $\sigma=2$. We can
 express the single-particle entanglement in another way $S_{1}\approx\ln[2(N/k-1)+1]$. It's known that for a Laughlin state of
  $N_{0}$ particles in spherical geometry with filling factor $\nu=1/m$, $S_{1}^{\textmd{Lau}}=\ln[m(N_{0}-1)+1]$.
  Therefore in the viewpoint of entanglement, when $N/k$ is an integer, our subspace ground state
behaves approximately like a Laughlin state of $N/k$ particles with
$m=2$, namely in our subspace ground state one particle is only
entangled with $N/k-1$ particles in the form of Laughlin state.
Enlightened by this fact, we can express the subspace ground state
$\Psi^{k}$ simply as a product of $k$ Laughlin states, each of which
consists of $N/k$ particles and has a filling factor 1/2. Then
considering the symmetry of boson wave function, we obtain: (the
exponential factor is omitted and we suppose $N$ is divisible by
$k$)
\begin{eqnarray}
\Psi^{k}=\mathcal {S}\Big[\prod_{i<j\in
A_{1}}^{N/k}(z_{i}-z_{j})^{2}...\prod_{l<m\in
A_{k}}^{N/k}(z_{l}-z_{m})^{2}\Big],\nonumber
\end{eqnarray}
%\begin{widetext}
%\begin{eqnarray}
%\Psi^{k}=\mathcal {S}\Big[\prod_{i<j\in A_{1}}^{\frac{N-\bar{N}}{k}+1}(z_{i}-z_{j})^{2}...\prod_{l<m\in A_{\bar{N}}}^{\frac{N-\bar{N}}{k}+1}(z_{l}-z_{m})^{2}
%\prod_{p<q\in B_{1}}^{\frac{N-\bar{N}}{k}}(z_{p}-z_{q})^{2}...\prod_{r<s\in B_{k-\bar{N}}}^{\frac{N-\bar{N}}{k}}(z_{r}-z_{s})^{2}\Big],\nonumber
%\end{eqnarray}
%\end{widetext}
where the set $A_{i}(i=1,2,...,k)$ has $N/k$ particles and the
symbol $\mathcal {S}$ indicates symmetrization over all partitions
of $N$ particles into sets $A_{i}$. Now we obtain the same trial
quantum Hall state as in Ref.\cite{NNJ} but we achieve this in an
entanglement manner. It has been checked that $\Psi^{k}$ has large
overlap with the ground state obtained by exact diagonalization
\cite{NNJ}. If $N/k$ is not an integer, namely $\bar{N}\neq0$, we
can slightly change the form of $\Psi^{k}$:
\begin{eqnarray}
\Psi^{k}=\mathcal
{S}\Big[\prod_{i<j\in
A_{1}}^{\frac{N-\bar{N}}{k}+1}(z_{i}-z_{j})^{2}...\prod_{l<m\in
A_{\bar{N}}}^{\frac{N-\bar{N}}{k}+1}(z_{l}-z_{m})^{2}\nonumber\\ \times
\prod_{p<q\in
B_{1}}^{\frac{N-\bar{N}}{k}}(z_{p}-z_{q})^{2}...\prod_{r<s\in
B_{k-\bar{N}}}^{\frac{N-\bar{N}}{k}}(z_{r}-z_{s})^{2}\Big],\nonumber
\end{eqnarray}
where the set $A_{i}(i=1,2,...,\bar{N})$ has $(N-\bar{N})/k+1$ particles
and the set $B_{i}(i=1,2,...,k-\bar{N})$ has $(N-\bar{N})/k$
particles. One can verify that the total angular momentum of this
state is just
$L=\bar{N}(\frac{N-\bar{N}}{k})(\frac{N-\bar{N}}{k}+1)+(k-\bar{N})(\frac{N-\bar{N}}{k})(\frac{N-\bar{N}}{k}-1)=(N-\bar{N})(N+\bar{N}-k)/k$.

\begin{figure}
\includegraphics[height=9cm,width=\linewidth]{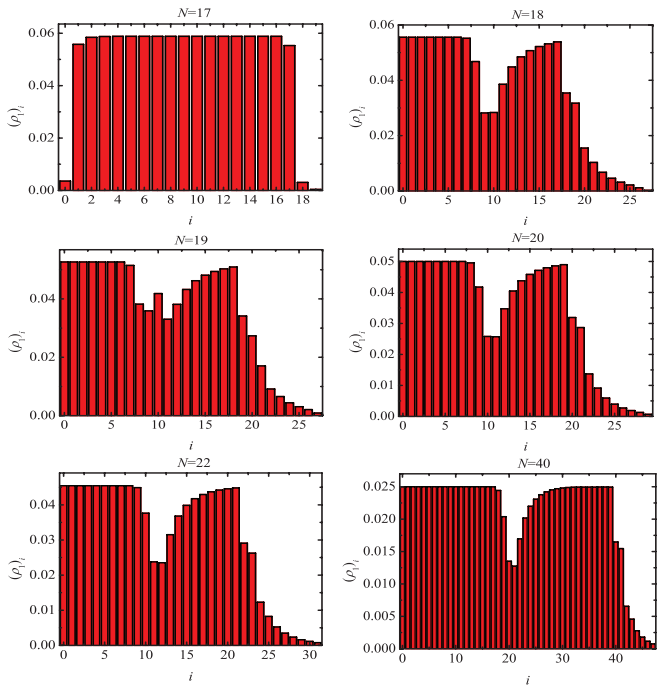}
\caption{\label{f5}(Color Online) The eigenvalues of the
single-particle reduced density operator of the ground state in the
subspace $\Delta L=N$ for fermions with repulsive Coulomb interaction. It
can be seen that at $N=18$, the distribution of eigenvalues changes
qualitatively.}
\end{figure}

\begin{figure}
\includegraphics[height=4cm,width=\linewidth]{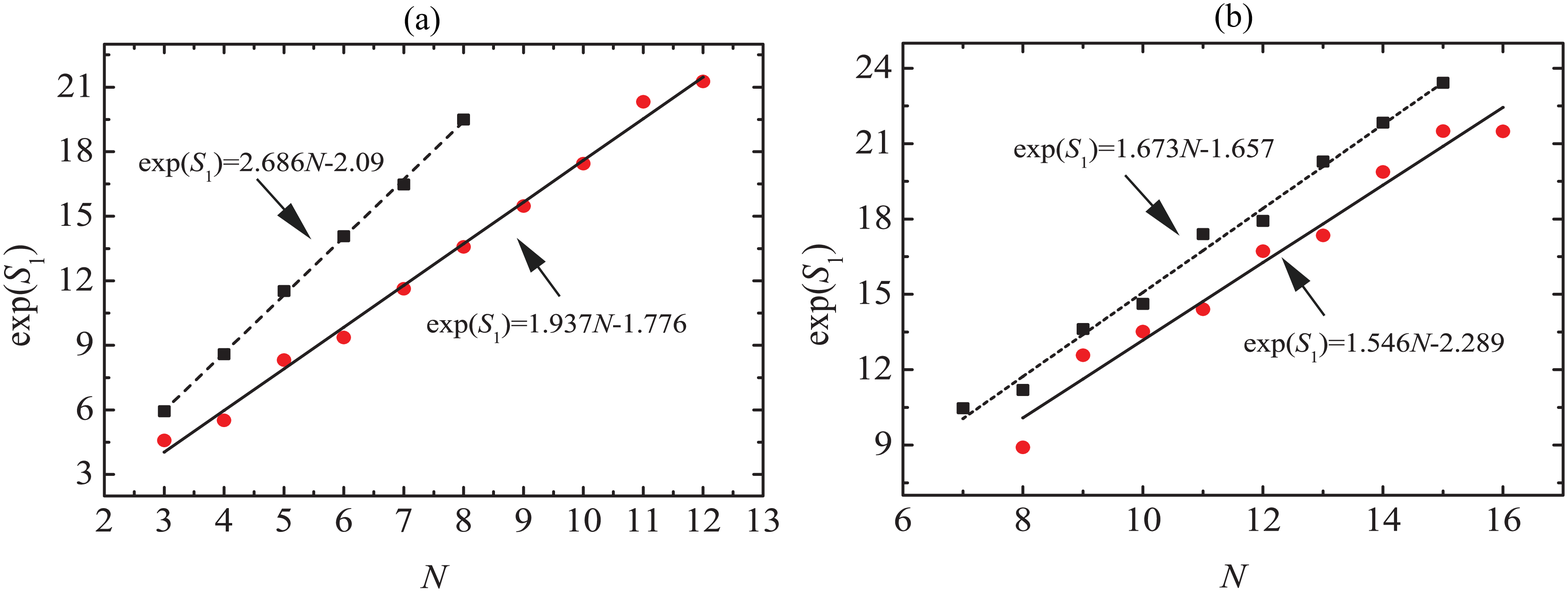}
\caption{\label{f6}(Color Online) Entanglement of fermions with
repulsive Coulomb interaction. (a) Black cubic: The entanglement of
the ground state in the subspace $\Delta L=N(N-1)$, namely $k=1$.
Red circle: The entanglement of the ground state in the subspace
$\Delta L=(N-1)^{2}/2$ for odd $N$ and $\Delta L=N(N-2)/2$ for even
$N$, namely $k=2$. (b) Black cubic: The entanglement of the ground
state in the subspace $\Delta L=(N-\bar{N})(N+\bar{N}-3)/3$, namely
$k=3$. Red circle: The entanglement of the ground state in the
subspace $ \Delta L=(N-\bar{N})(N+\bar{N}-4)/4$, namely $k=4$.}
\end{figure}

At last we study the single-particle entanglement of fermions of the
ground state in some special subspaces as a comparison with bosons.
The first subspace is $\Delta L=N$. In Fig.\ref{f3}, we plot $\Delta
S_{1}$ of the ground state in the subspace $\Delta L=N$ for
fermions. When $N$ is small, $\Delta S_{1}$ decays with $N$
monotonically similar to bosons. But when $N>17$, $\Delta S_{1}$
suddenly jumps to a higher position and then begins to decay again
accompanied by oscillation. This is because the distribution of the
eigenvalues of the single-particle reduce density operator changes
qualitatively from $N=17$ to $N=18$ (Fig.\ref{f5}). When
$N\leq17$, the subspace ground state is a central single-vortex
state, while when $N>17$, the subspace ground state is a candidate
of a off-center double-vortex state (see $N=18,20,22,40$ in Fig.\ref{f5}). Therefore
entanglement can indicate the edge reconstruction of rotating
fermions \cite{MT}. A similar phenomenon of electrons in quantum dot
that reconstruction of the maximum density droplet begins from the
edge rather than the dot center if the electron number exceeds
$N\approx15$ is reported in Ref.\cite{Yang}. The second subspace is
$\Delta L=(N-\bar{N})(N+\bar{N}-k)/k$. Through an analysis similar
to that for bosons, we can obtain that $S_{1}\approx\ln[(1+2/k)N-2]$
and find that the ground state in this subspace is a quantum Hall
state with filling factor $\nu=k/(k+2)$ and when $N/k$ is an integer
its counterpart in spherical geometry has a shift $\sigma=3$ (see Fig.\ref{f6}).

\section{Bosons in quadratic plus quartic trap}
So far we only consider the case in which the trap potential is a
harmonic potential $\mathcal {V}(r)=\frac{1}{2}m\omega^{2}r^{2}$.
This leads to the irrelevance of the subspace ground state with the
interaction strength $U_{0}$, because in Eq.(\ref{e2}) the subspace
ground state is uniquely determined by
$\sum_{i,j,k,l=0}^{L}U_{i,j,k,l}a_{i}^{\dag}a_{j}^{\dag}a_{k}a_{l}$.
Consequently, the subspace ground state entanglement is also
irrelevant of $U_{0}$. In this section, we consider a quadratic plus
quartic trap, namely
\begin{eqnarray}
V(r)=\frac{1}{2}m\omega^{2}r^{2}\Big[1+\lambda\Big(\frac{r}{a_{0}}\Big)^{2}\Big],\nonumber
\end{eqnarray}
where $a_{0}=\sqrt{\frac{\hbar}{m\omega}}$ is the oscillator
characteristic length. The rotating Bose and Fermi gas confined in
such an anharmonic potential is investigated in Refs.\cite{Ga,Gb}.
In the weakly-interacting limit, the subspace Hamiltonian can be
written as
\begin{eqnarray}
\mathcal
{H}_{L}=\sum_{l=0}^{L}\epsilon_{l}a_{l}^{\dag}a_{l}-L\hbar\Omega+U_{0}\sum_{i,j,k,l=0}^{L}U_{i,j,k,l}a_{i}^{\dag}a_{j}^{\dag}a_{k}a_{l},\label{e3}
\end{eqnarray}
where
\begin{eqnarray}
U_{i,j,k,l}=\int\!\!\!\!\int
dz_{1}dz_{2}\varphi_{i}^{*}(z_{1})\varphi_{j}^{*}(z_{2})\mathcal
{U}(z_{1},z_{2})\varphi_{k}(z_{2})\varphi_{l}(z_{1}).\nonumber
\end{eqnarray}
The single-particle LLL wave function $\varphi_{l}$ and its
eigen-energy $\epsilon_{l}$ are solved by Numerov method. The ground
state of Eq.(\ref{e3}) is dependent on $U_{0}$, making it possible
for us to investigate the relation between entanglement and
interaction. For simplicity we suppose that the bosons interact with
each other through contact potential (either attractive or
repulsive), so that $U_{i,j,k,l}=\int
dz\varphi_{i}^{*}(z)\varphi_{j}^{*}(z)\varphi_{k}(z)\varphi_{l}(z)$.
In order to keep the LLL approximation valid, throughout the
calculation we restrict the range of $U_{0}$ in the interval
$[-0.05\hbar\omega,0.05\hbar\omega]$.

\begin{figure}
\includegraphics[height=4cm,width=\linewidth]{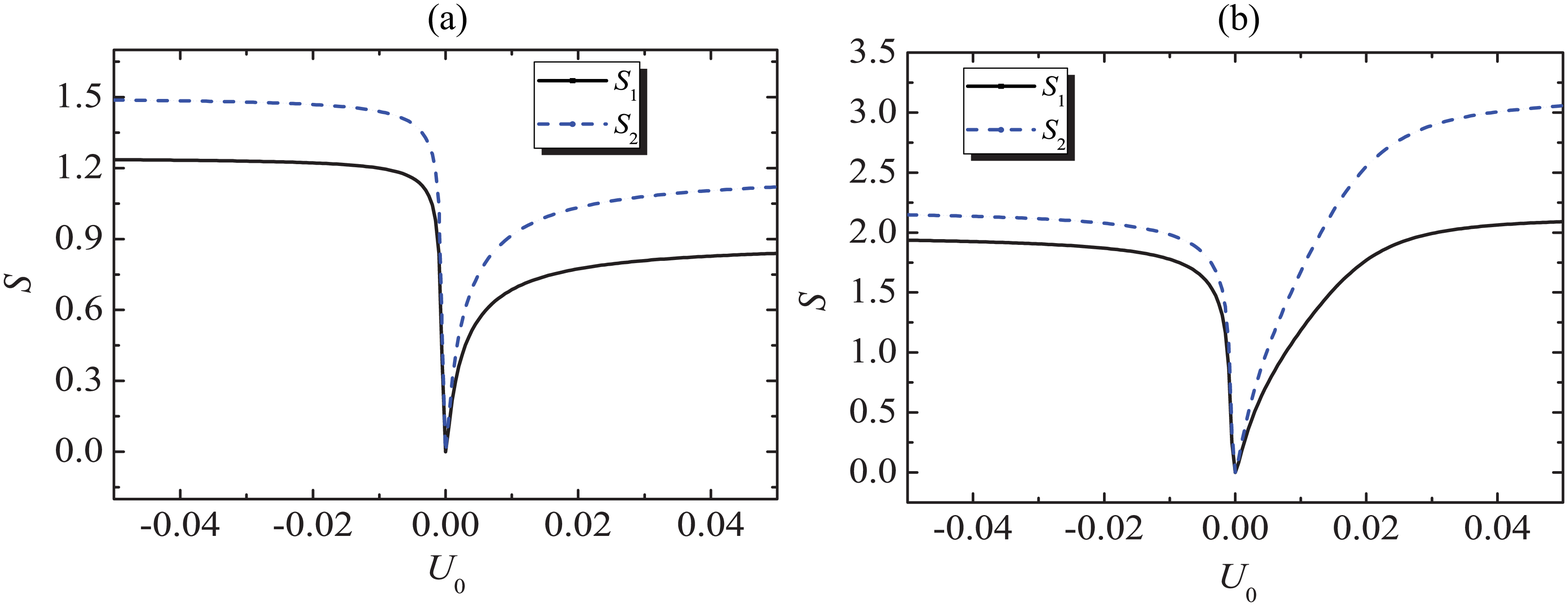}
\caption{\label{f7}(Color Online) Entanglement of bosons with
contact interaction, both attractive and repulsive ($\lambda=0.005$
in our calculation and $U_{0}$ in unit $\hbar\omega$). (a) The single-particle and two-particle
entanglement of the ground state in the subspace $L=N=5$. (b) The
single-particle and two-particle entanglement of the ground state in
the subspace $N=5$, $L=N(N-1)=20$.}
\end{figure}

\begin{figure}
\includegraphics[height=4cm,width=\linewidth]{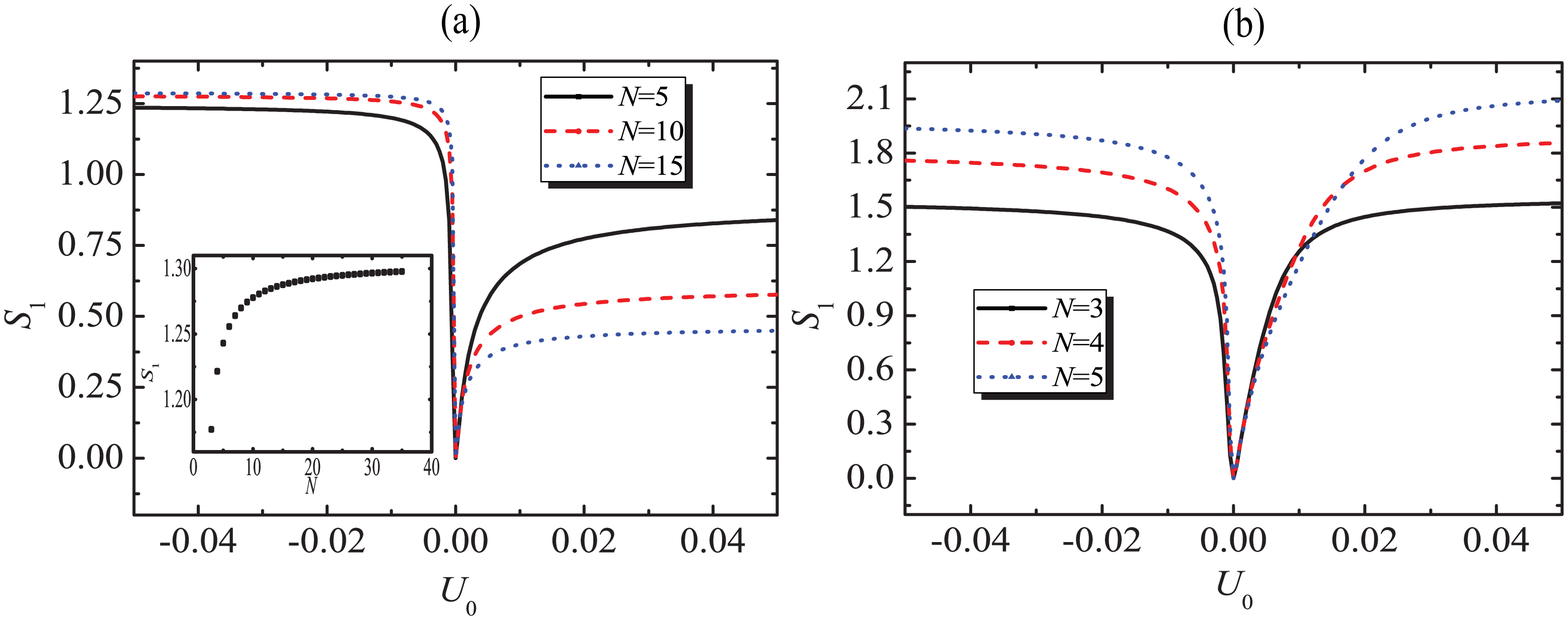}
\caption{\label{f8}(Color Online) Entanglement of bosons with
contact interaction, both attractive and repulsive ($\lambda=0.005$
in our calculation and $U_{0}$ in unit $\hbar\omega$). (a) The single-particle entanglement of the
ground state in the subspace $L=N$ for $N=5,10,15$. One can see that
when the interaction is negative, $S_{1}$ increases with $N$ while
for a positive interaction, $S_{1}$ decreases with $N$. (b) The
single-particle entanglement of the ground state in the subspace
$L=N(N-1)$ for $N=3,4,5$. The inset of (a) is the single-particle
entanglement in the subspace $L=N$ when the interaction is negative
with $\lambda=0$.}
\end{figure}

We consider both the single-particle and two-particle entanglement
of the ground state in the subspace $L=N$ and $L=N(N-1)$. From
Figs.\ref{f7} and \ref{f8}, we can see that both $S_{1}$ and $S_{2}$
grow with the increase of $|U_{0}|$, however, the curves are not
symmetric with respect to the point $U_{0}=0$, where no entanglement exists. In Fig.\ref{f8}, we
note that when the interaction is negative, the single-particle
entanglement of the ground state in the subspace $L=N$ increases
with $N$, meaning that attractive bosons cannot condense. However,
when the interaction is positive, the single-particle entanglement
of the ground state in the subspace $L=N$ decreases with $N$,
leading to a condensate phase in the thermodynamic limit.

\section{Summary}
In this paper, we consider the particle entanglement in rotating
Bose and Fermi gases. When the particle number $N$ is fixed, for
both bosons and fermions, through investigating the single-particle
entanglement of the ground state in various subspaces with fixed
angular momentum $L$, we find that the phenomena of particle
localization and vortex localization can be indicated by the
single-particle entanglement. Moreover, we study the single-particle
entanglement of the ground state in some special subspaces. For the
subspace $L=N$, through the comparison between the results of bosons
and fermions, we find that the edge reconstruction of rotating
fermions can be indicated by the single-particle entanglement. For
the subspace $L=(N-\bar{N})(N+\bar{N}-k)/k$, we relate the subspace
ground state with quantum Hall state from the viewpoint of
entanglement and construct the same trial wave function as that in
Ref.\cite{NNJ}. For bosons, different properties of condensate phase
and vortex liquid phase is reflected by entanglement. At last, we
study the relation between entanglement and interaction by
introducing a quartic trap.

\begin{acknowledgments}
Zhao Liu thanks Hongli Guo for the very important help in numerical
calculation. This work is supported by grants of National Natural
Science Foundation of China (NSFC) No.(10974247), ``973" program
(2010CB922904) of Ministry of Science and Technology (MOST), China.
\end{acknowledgments}

%\newpage
\bibliography{apssamp}
\end{document}